\documentstyle[aps,prl,floats,twocolumn,epsf]{revtex}

\def\gsim{\:\raisebox{-0.5ex}{$\stackrel{\textstyle>}{\sim}$}\:}
\begin{document}

\title{Nonequilibrium Fluctuations, Travelling Waves,
and Instabilities in Active Membranes} \author{Sriram
Ramaswamy$^1$, John Toner$^2$, and Jacques Prost$^3$\\
$^1$Centre for Condensed Matter Theory,
Department of Physics, Indian
  Institute of Science, Bangalore 560 012 INDIA\\
$^2$Material Science Institute, Institute of
Theoretical Science, and Department of Physics, University of
Oregon, Eugene OR 97403-5203 USA  \\
$^3$Institut Curie, Section de Recherche, 26 rue
d'Ulm, 75231 Paris cedex 05 FRANCE}
\date{\today}
\maketitle

\begin{abstract}
The stability of a flexible fluid membrane containing  a
distribution of mobile, active proteins (e.g., proton pumps)
is shown to depend on the structure and functional asymmetry
of the proteins. A {\em stable} active membrane is in a
nonequilibrium steady state with height fluctuations whose
statistical properties are governed by the protein activity.
Disturbances are predicted to travel as waves at sufficiently
long wavelength, with speed set by the normal velocity of the
pumps. The {\em unstable} case involves a spontaneous,
pump-driven undulation of the membrane, with clumping of the
proteins  in regions of high activity.
\end{abstract}

The functioning of active proteins in energy-dissipating
processes, such as ion transport, protein translocation,
and biopolymer synthesis, generates forces on the membranes
of the living cell and its organelles \cite{dlb,hous}.
As the active proteins diffuse around in
the membrane, the resulting fluctuations in
this force provide a nonthermal source of noise
for shape fluctuations of the membrane. The membranes
of a living cell are therefore nonequilibrium
or {\em active} membranes. Although such active,
nonequilibrium processes are abundant
in biological membranes, physicists have focussed mainly ---
with considerable success \cite{helfrich,udo} --- on
the statistical mechanics of membranes at thermal equilibrium.
There are however reasons \cite{levkor,tuvia} to suspect
that nonequilibrium processes are at work even in
red-blood-cell flicker, traditionally explained as
thermal equilibrium shape fluctuations \cite{bl}.
The predictions \cite{jprb,jpjbmrb} of fluctuation enhancement
in active membranes and the micropipette experiments
\cite{jbmpbdljp} on membranes laden with the photoactive proton pump
bacteriorhodopsin (bR) are further motivation for our studies.

\begin{figure}
\centerline{\epsfxsize=6cm \epsfysize=2cm \epsfbox{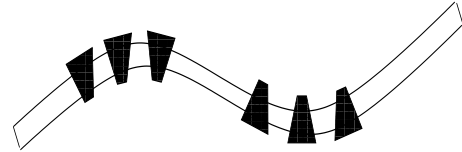}}
\caption[]{\label{conccurv} Asymmetric proteins (the black wedges)
imbedded in a fluid membrane are drawn to regions with curvature adapted
to the protein shape.}
\end{figure}

In this Letter, we consider the statistical mechanics
and dynamics of a fluid membrane containing a distribution of
identical, active pumps free to move in the plane of the membrane.
By ``pump'' we mean a membrane-spanning protein which,
when supplied with energy, transfers material (and thus exerts a force on the
membrane) in one direction only.  This ignores the complexities of some
real pumps but is a good description  of bR.
By convention, we shall term the end towards which the force acts the
{\em head} of the pump, and the other end its {\em tail}.
We shall call a protein a ``$+$ pump'' (``$-$ pump'') if the vector
from its tail to its head points {\em parallel to} ({\em antiparallel to})
a fixed outward normal of the membrane.
Flips from $+$ to $-$, prohibitively slow
in any real system, are forbidden in our treatment.
A membrane will be termed ``balanced'' (``unbalanced'')
if the numbers of $+$ and $-$ pumps are equal (unequal).
The unidirectional pumping action means that a given pump knows up
from down. In general, therefore, (i) a pump of a given sign will
favor one sign of the local mean-curvature $H$ of the membrane
over the other (Fig. \ref{conccurv}), and (ii) its pumping
activity will depend on $H$. Incorporating these effects
distinguishes the present work
from \cite{jprb,jpjbmrb}.

Here is a summary of our main results:
(i) An active membrane can be either linearly {\em stable} or
linearly {\em unstable} against undulation and
pump aggregation, depending on the structural and functional
asymmetries mentioned in the previous paragraph.
(ii) In the {\em stable} case, for plausible
values of the physical parameters
a ``balanced'' active membrane has,
over an appreciable intermediate range of wavenumbers $k$,
a height variance $\langle |h_k|^2\rangle$
$\propto 1/k^4$, with a coefficient
{\em independent} of the concentration of pumps.
These are truly nonequilibrium fluctuations, with
a strength depending on {\em kinetic} coefficients:
$\langle |h_k|^2\rangle$ is proportional to the permeability of
the membrane.   This is broadly consistent with the
observations of \cite{jbmpbdljp}.  (iii) For a {\em stable}
membrane, the longest wavelength disturbances travel as
non-dispersive waves, i.e., waves with a speed independent of
wavevector k.  The wave speed $c$ is set by the pump activity
and independent of the membrane elasticity.   (iv) In this
longest wavelength regime, the height variance is much
smaller, as $k \rightarrow 0$, than in the equilibrium case;
specifically $\langle \left|h_k\right|^2\rangle \propto {1
\over k^2}$ as $k \rightarrow 0$.  This result depends on the
effect of nonlinearities, which also determine the final
state of the system \cite{ashwin,sunil} in the unstable case.

We now construct our model and derive the results stated
above. We characterize the active membrane by its mean curvature
$H$ and the areal concentrations $\nu_{\pm} $ of $+$ and $-$
pumps .  In a Monge description $H \simeq -\nabla^2h$ and
$\nu_{\pm} = n_{\pm}/\sqrt{1 + (\nabla h)^2}$ where
$h({\bf x},t)$ is the height of the membrane at time $t$ above point
${\bf x}$ on a two-dimensional reference plane, and
$n_{\pm}({\bf x},t)$ are the {\em projections}
of the pump concentrations onto the plane.
It is useful to distinguish
the {\em protein concentration}
\begin{equation}
\label{totaldensity}
n({\bf x},t) \equiv n_+({\bf x},t) + n_-({\bf x},t)
\end{equation}
and the {\em signed protein density}
\begin{equation}
\label{signed}
m({\bf x},t) \equiv n_+({\bf x},t) - n_-({\bf x},t)
\end{equation}
with averages $m_0$ and $n_0$ respectively. For convenience, we shall
work with
\begin{equation}
\label{psiandphi}
\psi \equiv {m \over n_0}, \, \phi \equiv {n \over n_0}, \,
\psi_0 \equiv m_0 / n_0.
\end{equation}

We begin with a membrane {\em at thermal equilibrium}
at temperature $T \equiv \beta^{-1}$, where the proteins are {\em present}
but not {\em activated}. The probability of a configuration $\{h({\bf x}),
\psi({\bf x})\}$
is $\propto$ exp$(- \beta E)$ with a Hamiltonian \cite{andelman} 
\begin{eqnarray}
\label{curvenergy}
E\left[h,\psi\right] &=& {1 \over 2} \int \mbox{d}^2x [ \kappa (\nabla^2
h)^2 +
\sigma (\nabla h)^2
\nonumber \\
&&+ A(\psi - \psi_0)^2 - 2\kappa \bar{H} \psi \nabla^2h] 
\end{eqnarray}
to bilinear order in the variables involved.
$E$ contains a bending energy \cite{helfrich} with a rigidity $\kappa$,
a surface tension $\sigma$, a compression energy for the signed protein
density, with osmotic modulus
\begin{equation}
\label{osmod}
A \sim Tn_0
\end{equation}
for $n_0 \to 0$,
and a coupling $\bar{H}$ which determines the local spontaneous curvature
induced \cite{marc} by the presence of a protein (Fig. \ref{conccurv}).
This last term is one of the consequences of the directionality of
the pumps, and can
be attributed to a head-tail size difference of magnitude
\begin{equation}
\label{ell1}
\ell_1  \equiv {\bar{H} \over n_0 }.
\end{equation}
We expect $\ell_1$ to be independent of $n_0$ for $n_0 \to 0$.

We now explain why (\ref{curvenergy}) ignores the
protein concentration field $\phi$.
Minimizing (\ref{curvenergy}) tells us that in thermal
equilibrium a membrane with a net imbalance $\psi_0$ of pumps will
develop a spontaneous curvature
\begin{equation}
\label{psi0}
H_0 \equiv \left(\nabla^2h \right)_{\mbox{min E}}
= {\bar{H} \over 1 - {\kappa\bar{H}^2
\over A}}
\, \psi_0 \quad .
\end{equation}
Many experiments on the physics of membranes are carried
out on giant vesicles (size $\gsim 20 \mu$m), for which the mean curvature
and hence, from (\ref{psi0}), $\psi_0$ are negligibly small.  Accordingly,
we shall work at
$\psi_0~=~0$ for much of this paper.  In this limit, the
symmetry $h \to -h, \, \psi \to -\psi \, (\mbox{equivalently} \,
\epsilon \to -\epsilon$ for each pump of sign $\epsilon = \pm$)
rules out any bilinear coupling $\psi \phi$ in (\ref{curvenergy})
for $\psi_0 = 0$. Thus, in a linearized treatment of a
balanced membrane, $\phi$ decouples from $h$ and $\psi$.
In addition, for $\psi_0 = 0$,
the mean normal drift speed of the membrane in the active
state is zero. Towards the end of this paper we shall
present some important results for $\psi_0 \neq 0$.

Equation (\ref{curvenergy})
implies the {\em equilibrium}
height variance
 \begin{equation}
\label{hvarequil}
\langle |h_k|^2 \rangle = {T \over {\sigma k^2 + \kappa_{eff} k^4}}
\end{equation}
at wavevector ${\bf k}$, with an effective rigidity
\begin{equation}
\label{kappaeff}
\kappa_{eff} = \kappa - {(\kappa \bar{H})^2 \over A}
\end{equation}
independent of sgn$(\bar{H})$.
The dynamics of small fluctuations is
also determined by $\kappa_{eff}$. We shall assume that  
$\bar{H}$ is
small
enough to keep $\kappa_{eff} > 0$ \cite{footnote1},
so that the presence
of the inactive but shape-asymmetric
proteins merely shifts the value of the
bending rigidity. 

In the active state (e.g., when the bR in the experiments of
\cite{jbmpbdljp}  is illuminated with green light) even static
quantities such as
$\langle |h_k|^2 \rangle$ must be determined from the {\em dynamical}
properties of the system. To this end, let us assemble the ingredients
for the equations of motion, to leading orders in a gradient
expansion, of a membrane with active proteins.
We continue to work at $\psi_0 = 0$.

An isolated active pump with sign $\epsilon = \pm$ exerts a force
$\epsilon F_a$ normal to the membrane, if the local mean curvature
$H = 0$.  If $H \neq 0$, the symmetry
$h \to -h, \, \epsilon \to -\epsilon$ permits
an additional contribution $\ell_2 H F_a$ to the force, where $\ell_2$,
whose sign is not fixed by symmetry, is a length characterizing
the sensitivity of the pumping mechanism to the bending
of the membrane.
The force arising from the activity of a distribution
of $+$ and $-$ pumps with intrinsic concentrations $\nu_{\pm}$,
together with the elastic force $f_{el}$ arising from
(\ref{curvenergy}), give rise via permeative flow
with kinetic coefficient $\mu_p$ \cite{permeation} to a normal velocity
\begin{eqnarray}
\label{vnormal}
v_n &=& \mu_p[(\nu_+ - \nu_-) + (\nu_+ + \nu_-)\ell_2 H \nonumber\\
&&+ O(\nu_+ - \nu_-)^3] F_a - \mu_p f_{el},
\end{eqnarray}
where the $O(\nu_+ - \nu_-)^3$ (and higher odd order) terms arise only
if the presence of a given pump affects the activity of other pumps.

Projecting (\ref{vnormal}) {\em normal} to the reference
horizontal plane \cite{cai},
defining the natural velocity scale
\begin{equation}
\label{v0}
v_0 \equiv \mu_p F_a n_0 > 0,
\end{equation}
adding the contribution $v_{\tiny hyd}$ due to {\em
hydrodynamic} flow \cite{bl,intrinsic}
and expanding in powers of $\nabla h$, we obtain
\begin{eqnarray}
\label{heqn}
\frac{\partial h}{\partial t} & = &
v_0 [\psi + \ell_2 \phi \nabla^2 h + O(\psi^3, \psi^3 (\nabla h)^2)]
\nonumber \\
&&- \mu_p \delta E / \delta h
+ v_{\tiny hyd} + f_h
 \nonumber \\
& \simeq & v_0 (\psi + \ell_2 \nabla^2 h) + v_{\tiny hyd} 
\nonumber \\
&&-  \mu_p (-\sigma \nabla^2 h + \kappa \nabla^4 h 
- \kappa \bar{H} \nabla^2 \psi) 
+ f_h.
\end{eqnarray}
Here the Gaussian zero-mean noise $f_h$ has variance \cite{ymsr,jpjbmrb}
$2 T [\mu_p + (1 /4 \eta q)^{-1}]$ at wavenumber $q$, and
\begin{equation}
\label{vhyd}
v_{\tiny hyd} = -\int {\mbox{d}^2 q \over (2 \pi)^2} e^{i{\bf q}.{\bf x}}
{1 \over {4 \eta q}} {\delta E \over \delta h_{\bf q}(t)},
\end{equation}
where $h_{\bf q}(t)$ is the spatially Fourier-transformed height field and
$\eta$ the
solvent viscosity.
The signed protein density obeys the conservation law \cite{noq3forpsi}
\begin{eqnarray}
\label{psieqn}
\frac{\partial \psi}{\partial t} & = & v_0 \nabla \cdot
\left({\psi^2
\nabla h \over {1 + (\nabla h)^2}}\right)
+ \Lambda \nabla^2 \delta E / \delta \psi + \nabla \cdot {\bf
f}_{\psi}
\nonumber \\
& \simeq & \Lambda A \nabla^2 \psi - \Lambda \kappa \bar{H}
\nabla^4 h + \nabla \cdot {\bf f}_{\psi},
\end{eqnarray}
where the $v_0$ term comes from projecting (\ref{vnormal}) {\em parallel}
to the reference plane, $\Lambda$ is the mobility
of the pumps,
and ${\bf f}_{\psi}$ is a spatiotemporally
white, isotropic, vector thermal noise with variance $2 T \Lambda$.
Since the dissipative processes that gave rise to $f_h$
and ${\bf f}_{\psi}$ are still present, we ignore for
simplicity possible additional, pumping-dependent
bare noise sources.
In both (\ref{heqn}) and (\ref{psieqn}), the first equality
is valid for a general energy function $E$ including $\phi \psi$ couplings,
while the second (approximate) equality applies only to a {\em strictly}
linearized treatment at $\psi_0 = 0$
with $\phi = 1$. Equations (\ref{heqn}) and (\ref{psieqn}) define our model.

\begin{figure}
\centerline{\epsfxsize=6cm \epsfysize=4cm \epsfbox{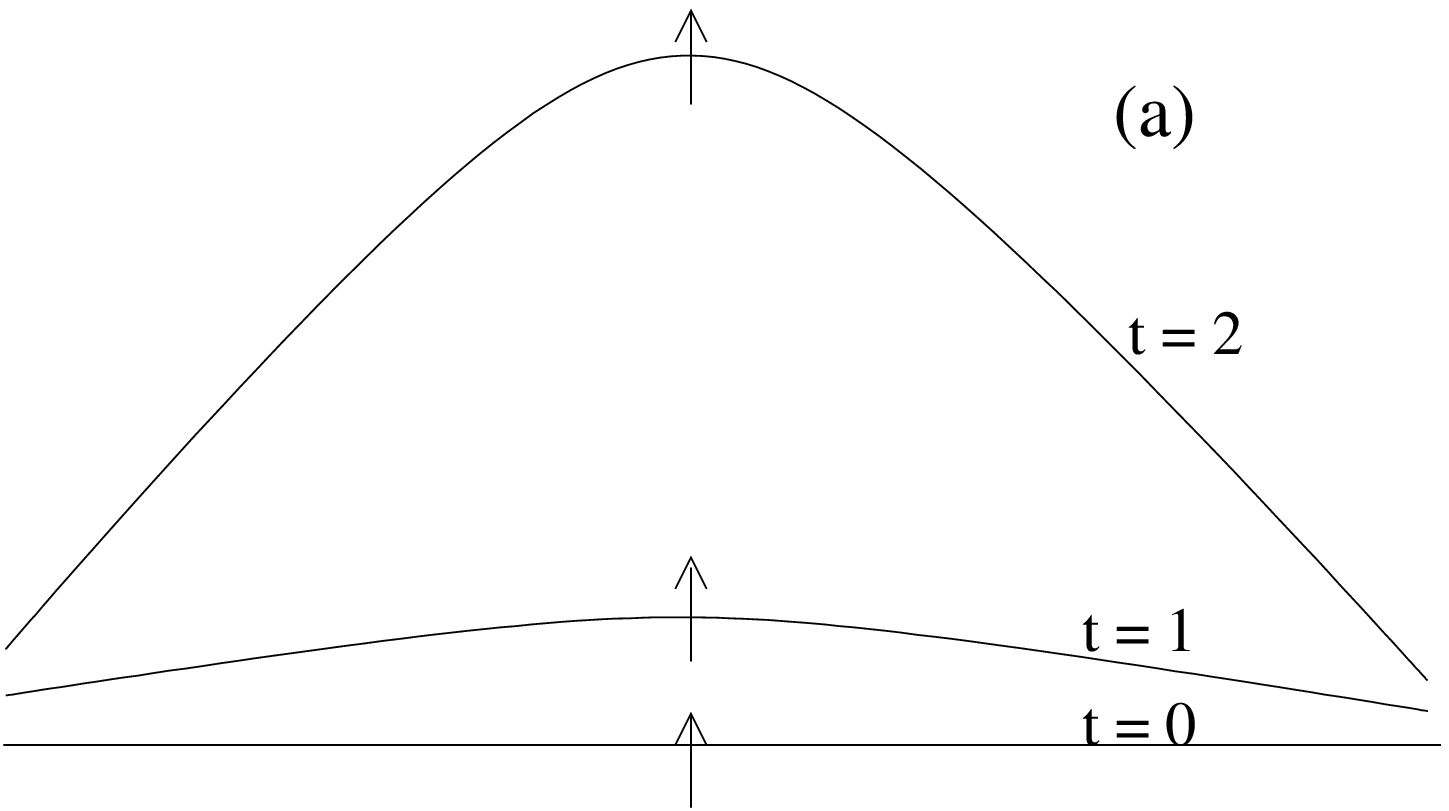}}
\centerline{\epsfxsize=6cm \epsfysize=6cm \epsfbox{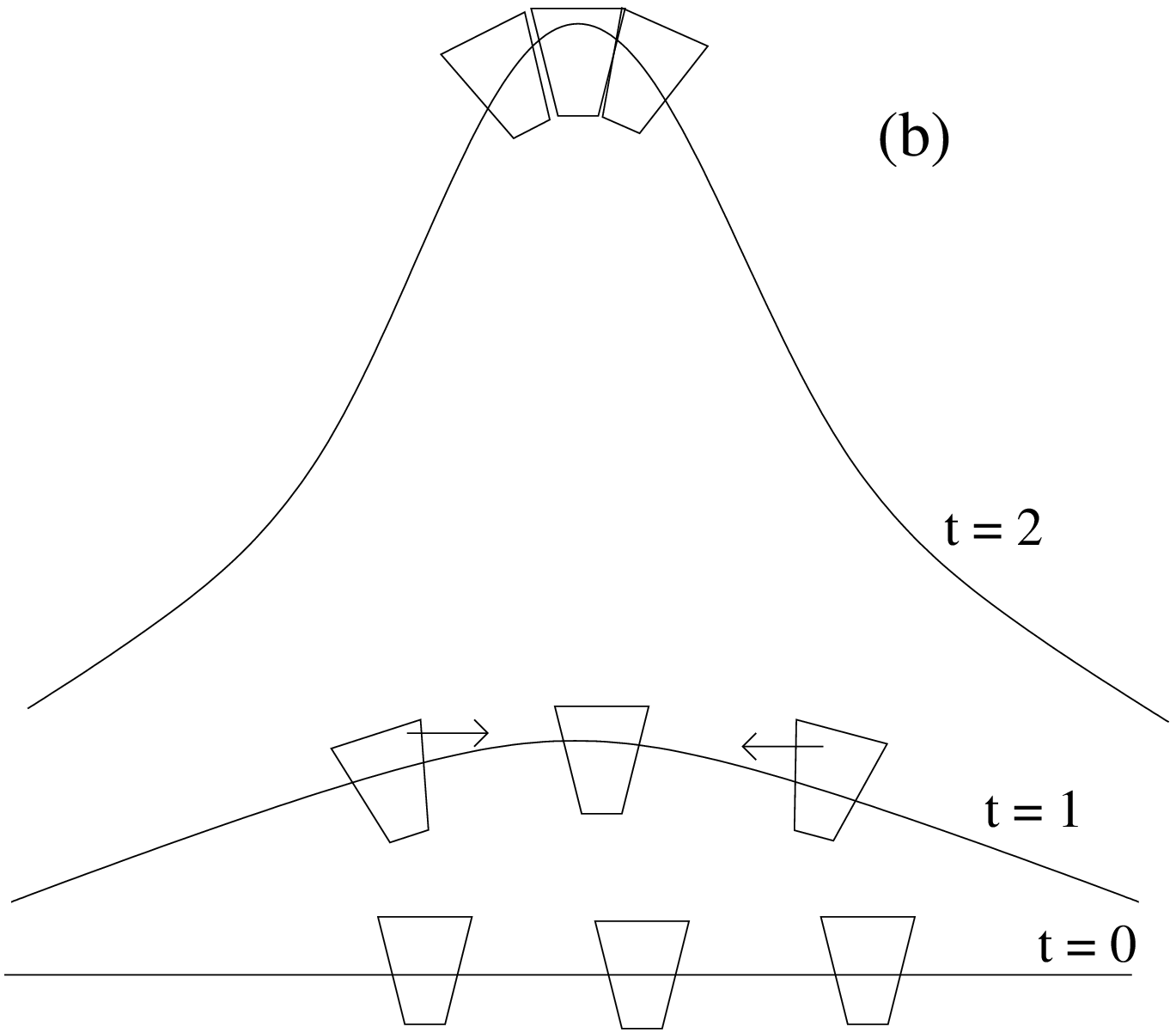}}
\caption[]{\label{instabfig} Two mechanisms for instability:
(a) If downward curvature enhances the upward force exerted by
a single pump, and vice versa; (b) if the curvature produced
by pumping attracts more pumps.}
\end{figure}

Before using these equations, it is crucial to note that
if $\ell_2$ or $\bar{H}$ is sufficiently large and negative,
then (\ref{heqn}) and (\ref{psieqn}) predict that a flat membrane
with a statistically uniform distribution of $+$ and $-$
pumps is linearly unstable, even if $\kappa_{eff} > 0$
(see eqn. (\ref{kappaeff})), with perturbations growing
at a rate proportional to $k^2$ at small wavenumber $k$. To
see this physically (Fig. \ref{instabfig}),
imagine a region
of excess active
$+$ pumps surrounded by a
statistically uniform $+ -$ mixture, in an initially
flat membrane. Pumping will pull the $+$ region
ahead of its surroundings leading to mean curvature
$H = -\nabla^2 h > 0$ around the $+$ pump. This
can lead to an instability in two ways:
(i) If $\bar{H} < 0$, this will attract more
$+$ pumps. (ii) if $\ell_2 < 0$, the
activity of the $+$ pumps in the region is
enhanced.
Such physical mechanisms could well be involved
in processes where cell membranes undergo large
deformations.
Well beyond the onset of these instabilities, nonlinearities
will determine the ultimate fate of the membrane \cite{ashwin,sunil,unpub}.

For the remainder of this paper, we assume parameter
values corresponding to a {\em linearly stable} active membrane.
The two eigenmodes for disturbances at small wavenumber $k$
are diffusive if $v_0 \ell_1, \, v_0 \ell_2$ are small
compared to the pump diffusivity $\Lambda A$,
and propagative but highly dispersive (wavespeed $\sim k$) if
$v_0$ is large enough.

More important is the
the height variance
$\langle |h_k|^2 \rangle$
at wavenumber $k$,
which is obtained  by linearizing and Fourier transforming
(\ref{heqn}) and (\ref{psieqn}) in space and time, solving for
$h_{{\bf k}, \omega}$
and $\psi_{{\bf k}, \omega}$ at wavevector $\bf k$ and frequency
$\omega$, using the statistical properties of the noise sources $f_h$ and
${\bf f}_\psi$ to obtain
$\langle |h_{{\bf k}, \omega}|^2 \rangle$, and integrating over $\omega$.

For $k$ much smaller than the lesser
of
\begin{equation}
\label{kmaxetc}
k_{max} \equiv {4 \eta v_0 \bar{H} \over A} \, \mbox{and} \,
k_D \equiv {4 \eta D_{\psi} \over \kappa},
\end{equation}
setting the tension $\sigma =0$,
we find from (\ref{heqn}) and (\ref{psieqn}) that
\begin{equation}
\label{hk2}
\langle |h_k|^2 \rangle =
{T \over {F_a n_0 \bar{\ell} k^2 + \kappa_{eff} k^4}} + \frac{\mu_p
F_a}{D_{\psi} \bar{\ell} k^4}
\end{equation}
where we have defined $\bar{\ell} \equiv \ell_2 + {\kappa n_0
\over A} \ell_1$.  The thermal wandering of the
membrane is thus {\em suppressed} by a pumping-induced
tension $F_a n_0
\bar{\ell}$, but a novel {\em nonequilibrium}  contribution
to the height fluctuations now appears, mimicking a
zero-tension  equilibrium membrane. Note that the coupling
$\bar{\ell}$ of the pumps to the  curvature is crucial here: for
$\bar{\ell} = 0$ we would find $k^{-5}$  behavior as in \cite{jprb}.
For $n_0 \to 0$, the diffusivity $D_{\psi} \equiv \Lambda A$  of the
pumps approaches that of an isolated pump and is hence nonzero and,
from (\ref{osmod}) and (\ref{ell1}), the  length
$\bar{\ell}  \simeq \ell_2 + (\kappa / T)
\ell_1$.  The coefficient of $k^{-4}$ in (\ref{hk2}) is thus
independent of $n_0$ for small $n_0$.

If the tension is nonzero, the behavior in (\ref{hk2}) is cut off
for $k$ less than
\begin{equation}
\label{kmin}
k_{min} \equiv {\sigma A \over 4 \eta v_0 \kappa \bar{H}} = {\sigma \over
\kappa k_{max}}.
\end{equation}
As a consequence, the excess area \cite{evans}
\begin{equation}
\label{excess}
\alpha \simeq {1 \over 2} \langle (\nabla h)^2 \rangle \simeq  \frac{\mu_p
F_a}{D_{\psi}\bar{\ell}} \ln{k_{max} \over k_{min}}
\end{equation}
can be seen to depend only logarithmically on the protein density $n_0$.
Moreover, increasing the solvent viscosity should decrease
$\mu_p$ while leaving $D_{\psi}$ unaffected, and should
thus decrease $\alpha$. These predictions of (\ref{excess}) are
consistent with the experiments of \cite{jbmpbdljp}. However, not
enough is known about the values of parameters such as
$\mu_p, \, F_a, \, \mbox{and}\, \bar{\ell}$ to make a detailed comparison.
If instead $D_{\psi}$ is decreased, $\langle
|h_k|^2 \rangle$
should increase. Such predictions of the dependence of static correlations
on {\em kinetic} quantities allow clear tests of our model and
of the truly nonequilibrium nature of the fluctuations.

We turn finally to the wavelike modes mentioned at the start of
the paper.  A net excess of pumps of one sign ($\psi_0 \neq 0$) would yield
a term $v_0 \psi_0^2 \nabla^2 h$ in the linearized (\ref{psieqn}).
With (\ref{heqn}) this would lead to propagating
waves  with a speed $v_0 \psi_0 = \mu_p F_a m_0$. Even for
$\psi_0 = 0$, the presence of fluctuations means that $\langle
\psi^2 \rangle \neq 0$. Hence, improving upon our
strictly linearized treatment by letting $\psi^2 \to \langle \psi^2 \rangle$
in (\ref{psieqn}) leads to a prediction of waves with
speed $c \equiv v_0 \langle \psi^2 \rangle^{1/2} =
\mu_p F_a \langle m^2 \rangle^{1/2}$, and to a nonequilibrium
height variance
$\langle |h_k|^2 \rangle \sim (\mu_p v_0^2 T)/(v_0 \ell_2 +
D_{\psi})c^2 k^2$.  It is important to note that these waves
only appear for wavevectors $k < k_{\mbox{\tiny fluc}}
\equiv \sqrt{{v_0 \left< \psi^2 \right>\over D_{\psi} \ell
_2}}$; for
$k \gg k_{\mbox{\tiny fluc}}$ our earlier linearized results
apply.  Hence, for a system with small fluctuations
$\left<\psi^2\right>$, the linearized results (\ref{hk2})
and (\ref{excess}) could well hold
at the experimental wavenumbers $k$.

In any case, for $\psi_0 \neq 0$, a complete analysis requires,
even at a linearized
level, the inclusion of fluctuations in the protein density
$n$.  Replacing $\psi^2$ by its average, too, is simply a first
step in a complete treatment  of nonlinear fluctuation effects. A
detailed consideration of all such effects will appear elsewhere
\cite{unpub}.

We close by summarizing this work.
Starting with a natural model of a membrane with
active proteins, we show that
such a membrane can be linearly stable
or unstable to small perturbations in its shape
and the distribution of proteins. In the stable
case, we show that the $k^{-4}$ dependence
of the height fluctuations can mimic that of an
equilibrium membrane, but with an enhanced temperature,
consistent with the observations of \cite{jbmpbdljp}.
Further experiments, in particular on the
dependence of the height variance on the protein
diffusivity, will provide more stringent tests of our predictions.
We are currently in the process of studying the
effects of nonlinearities on the scaling of correlation
functions in the stable case and on the growth in the
unstable case\cite{ashwin,sunil,unpub}.

We thank F. J\"{u}licher, P.B.S. Kumar, R. Nityananda, A. Pande, 
R. Pandit, and M. Rao for valuable discussions,
the U.\ S.\ NSF (Grant Number DMR-9634596) for financial
support, and the Aspen Center for Physics and the Raman
Research Institute for their hospitality.


\begin{references}
\bibitem{noq3forpsi} The diffusive damping of $\psi$
is unaffected at small wavenumber by coupling to solvent flow in an
effective linearized treatment at $\psi_0 = 0 \cite{unpub}$.
\end{references}
\end{document}